# Requirements and Recommendations for IoT/IIoT Models to automate Security Assurance through Threat Modelling, Security Analysis and Penetration Testing

Ralph Ankele
ralph.ankele@joanneum.at
JOANNEUM RESEARCH Forschungsgesellschaft mbH
Graz, Austria

Stefan Marksteiner
stefan.marksteiner@avl.com
AVL List GmbH
Graz, Austria

Kai Nahrgang
kai.nahrgang@joanneum.at
JOANNEUM RESEARCH Forschungsgesellschaft mbH
Graz, Austria

Heribert Vallant
heribert.vallant@joanneum.at
JOANNEUM RESEARCH Forschungsgesellschaft mbH
Graz, Austria

## ABSTRACT
The factories of the future require efficient interconnection of their physical machines into the cyber space to cope with the emerging need of an increased uptime of machines, higher performance rates, an improved level of productivity and a collective collaboration along the supply chain. With the rapid growth of the Internet of Things (IoT), and its application in industrial areas, the so called Industrial Internet of Things (IIoT)/Industry 4.0 emerged. However, further to the rapid growth of IoT/IIoT systems, cyber attacks are an emerging threat and simple manual security testing can often not cope with the scale of large IoT/IIoT networks.

In this paper, we suggest to extract metadata from commonly used diagrams and models in a typical software development process, to automate the process of threat modelling, security analysis and penetration testing, without detailed prior security knowledge. In that context, we present requirements and recommendations for metadata in IoT/IIoT models that are needed as necessary input parameters of security assurance tools.

## KEYWORDS
IoT, IIoT, Industry 4.0, Penetration Testing, Threat Modelling, Security Analysis, Automation, Requirements, Recommendations

## 1 INTRODUCTION
With the Internet of Things (IoT) reaching manufacturing systems, the Industrial Internet of Things (IIoT) or Industry 4.0 was forged (see Figure 1) [10, 36]. This technology enhances traditional control systems to form Cyber-Physical Systems (CPS), a new class of systems that embeds cyber capabilities into the physical world. While this enables advanced reconfiguration and scalability, it however also introduces complexity and potential instability [42]. Enhancements for production include [57]:

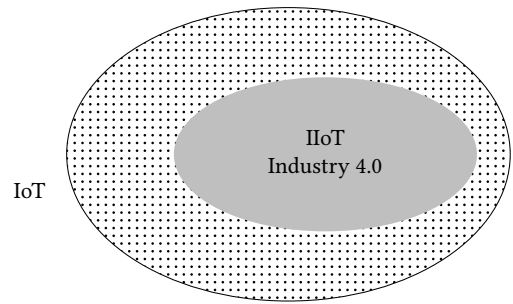

Figure 1: The Industrial Internet of Things (IIoT) is a subset of the Internet of Things (IoT).

- Horizontal integration through value networks;
- Vertical integration and networked manufacturing systems;
- End-to-end digital integration of engineering across the entire value chain.

These traits make the IIoT more useful and leverage proliferation – the expected growth of IoT in general is exponential (see Figure 2). This, however, also raises the attractiveness for adversaries to perform malicious actions to gain fame or profit [24]. Furthermore, connecting production networks both to each other and the Internet has consequences that are hard to estimate from a security perspective. This is because they can cause side effects. For instance, they could open an attack surface by introducing unintended connections and/or expose devices that are not meant to be connected and therefore lack of security features such as proper authentication to threats from external sources [49]. Also, many IoT protocols, both wired [59] and wireless [38], have known flaws that could allow for adversaries to launch successful attacks against a network. To counter this potential attacks, a structured approach to facilitate security assurance is needed, which involves design, analysis and testing. This paper therefore outlines building blocks for a systematic methodology of providing security assurance, specialized for the IIoT.

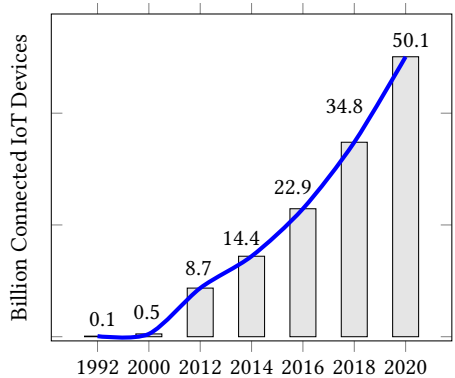

Figure 2: Estimated growth of connected devices in the IoT[1].

## 1.1 Research Gap and Related Work

With the rapid growth of the IoT (see Figure 2) also concerns have been raised that without appropriate considerations many new security challenges arise. In particular, as the IoT is interconnected and spreads widely, cyber attacks are an emerging threat [9, 31, 39, 47]. Therefore, researchers from academia and industry are analysing the security aspects of IoT systems [48]. Current research directions are mainly focusing on anomaly detection for security components in IoT systems [16, 46, 60] and penetration testing for IoT systems [12]. However, with the rapid growth of the IoT/IIoT systems, security analysis and manual security testing of an IoT system can often not cope with large scale IoT networks. Recent directions in security analysis of IoT systems have focused on automated security analysis. In most of the works so far, initially a detailed threat model of the IoT system under test has to be generated [58], then automated security tests [37] and penetration tests [3] are executed.

While this approach enables to automate penetration testing, still a detailed threat model has to be available. A natural further step would be to extract the necessary information needed to automate penetration tests already from more commonly used diagrams and models in the software development process (i.e. Unified Modelling Language, ...).

## 1.2 Contributions and Outline

In this work, we define requirements and recommendations of IoT/IIoT system models to include certain metadata that includes necessary input parameters of security assurance tools. This approach can further help to automate the process of penetration testing, threat modelling and security analysis, without any detailed security knowledge.

The remainder of this paper is structured as follows: In Section 2 we analyse the typical components of IoT and IIoT models and specify architectural patterns used in cyber-physical systems. Section 3 gives an overview of a typical security assurance process. Moreover, we review state-of-the-art penetration testing tools, threat modelling-tools, and give details on the security analysis process. Furthermore, we extract the required input parameter of most commonly used security assurance tools. In Section 4 we provide requirements and recommendations for the metadata used in IoT/IIoT models that need to be extracted to automate those security assurance tools. Finally, Section 5 concludes this work and states some future research questions.

[1]Data obtained from: https://www.comptia.org/resources/internet-of-things-insights-and-opportunities, Date accessed: 14.05.2019

## 2 ARCHITECTURAL PATTERNS AND CONCEPTS FOR CYBER-PHYSICAL SYSTEMS

Future factories require efficient interconnection of their physical machines to improve the level of productivity, performance, resource consumption, and uptime of machines with the purpose of generally increasing the collective collaboration along the supply chain. Higher production quality, reduced downtime and real-time competitive edge is what manufacturers want to achieve in an IIoT setting, compared to the currently existing setting of disconnected machines [11]. The basic technology platform that connects IoT and IIoT systems is also often referred to as Cyber-Physical System (CPS) [32, 33, 44].

Usually security testing is limited by resources (i.e. time, complexity of the system under test, etc.). To overcome these limitations, an actual IoT/IIoT system can be abstracted as a model. The model is an abstraction of the real-world IoT/IIoT system under test, that should optimally re-enact the expected behaviour of the system. In the following, we give a classification of the most commonly used devices and components in IoT/IIoT systems.

### 2.1 Smart Devices and Sensors

In IoT systems, these are all end-user devices and sensors that are used to collect information. In IIoT systems, these are a constituent part of cyber-physical systems (CPS), which integrate physical systems to software and communication systems. These devices include, for example, temperature sensors, thermostats, pressure sensors, humidity and moisture level sensors, light intensity detectors, proximity detection, RFID, and others.

### 2.2 Networks and Protocols

The IoT requires huge scalability in the network space to handle the vast number of interconnected devices. With billions of devices being added, networks and protocols have to be adapted to handle the data flow. While traditional networks like IPv4 Ethernet and WIFI are not optimised for IoT use cases [51] (e.g. the address space of IPv4 allows $2^{32}$ = 4.3 billion devices, while estimates show that there will be around 50 billion IoT devices in 2020 [53]), different networks and protocols emerged like Bluetooth LE [43], ZigBee [2], MQTT [7], Z-Wave [29], LoRAWAN [13], IPv6 [15], and others.

### 2.3 Gateways

In IoT networks, many different networks and protocols have to interact in association with each other. IoT gateways can be configured to manage bidirectional data traffic to ensure interconnectivity of devices and sensors as well as to ensure compatibility of network protocols.

## 2.4 Cloud/Servers

The IoT generates massive amounts of data from devices, sensors, applications and users that have to be managed efficiently. IIoT networks may require real-time analysis of sensor data and accurate analytics that cannot be provided by sensors due to limited resources and are, therefore, performed at different levels (edge, fog, cloud). Clouds are often further used to provide data storage, analytics, infrastructure and services to billions of interconnected devices.

## 2.5 Analysis/Actuators

Based on sensor data, several actuators in IoT networks can perform services and further cloud-based services can produce real-time insights. Big enterprises collect vast amounts of data that needs to be carefully analysed according to individual business use cases. In IIoT networks often several sensors simultaneously provide information that needs to be combined and analysed.

## 2.6 User Interface.

User interfaces are the accessible parts of the IoT that connect the user with devices. In IIoT networks the user interfaces are often replaced by well-defined APIs that offer an interface for other smart services and back-end systems, often referred to as machine-to-machine (M2M) communications.

## 2.7 Smart Services and Back-end Systems

While IoT systems tend to outsource their whole infrastructure, computation and storage to cloud services, IIoT systems mainly process information in back-end systems within the internal network of a company. Therefore, the interplay between cloud-based services and smart services in the back-end of a company has to be enabled.

## 2.8 Big Data Analytics/Artificial Intelligence/Machine Learning

The sensors and devices connected to IoT and IIoT networks produce vast amounts of data that need to be efficiently processed and analysed. Artificial intelligence (AI) is a field of computer science that describes machines that mimic cognitive functions of humans such as learning and problem solving. Machine learning (ML) is a core part of AI that allows software to predict outcomes and to find patterns without being explicitly programmed. Both AI and ML techniques are used in IoT/IIoT networks to efficiently analyse data [19, 30].

# 3 SECURITY ASSURANCE BY VIRTUE OF THREAT MODELLING, SECURITY ANALYSIS AND PENETRATION TESTING

When evaluating the security of IoT systems with certain security assurance methodologies and penetration testing tools, we have to define an attacker model, the procedure of penetration tests, as well as the information needed for the input parameters of the security assurance tools. A structured approach to define such a security development process is to follow a five steps development process and define security actions within each. Table 1 shows the mapping of security actions to the development process as outlined by the Security Development Lifecycle (SDL) [35].

Table 1: Development Phases and Security Actions [35].

| Development Phase | Security Action |
| --- | --- |
| Requirement | Inception |
| Design | Design & Threat Modelling |
| Implementation | Guidelines & Best Practices |
| Verification | Security Push (including penetration testing) |
| Release | Final Security Review |

The inception phase consists of project planning, while the final security review involves reviews (threat models), finishing (penetration testing) and documentation archiving. While this approach is intended for use with software, the process is similar for other purposes (for instance a cyber security implementation process for nuclear facility safety systems [41]). In order to provide means for security assurance of (I)IoT models, we concentrate on the design, implementation and verification phases of the SDL approach and specify the three main action groups in more detail:

- Threat modelling;
- Security analysis (to generate guidelines);
- Penetration testing (also called attack testing).

This can also be matched with the perspective of another model that describes the security needs as overlap of threats, design and goals [56]. The design part in that model is seen a little differently than in the SDL, as it is defined as features that define security and efforts to shape these features, where the latter resemble the guidelines generation in the SDL. Therefore, the security analysis in this paper can be seen as means to address the design, the threat modelling copes with the threat part of that model, and penetration testing is used to verify the fulfilment of security goals.

*Attacker Model.* We also have to consider different attacker models, to represent the capabilities of an attacker. For simplification, we consider the Dolev-Yao attacker model [18], which means that an attacker can overhear, intercept and synthesize any message in the network, and is only limited by the cryptographic methods that are used. Moreover, an attacker can encrypt and decrypt with any keys she knows/successfully obtains. For simplicity, we do not distinguish between remote attackers and attackers with physical access to devices in this work.

## 3.1 Threat Modelling

Threat modelling is an approach with the goal to identify threats and vulnerabilities within IT system architectures [52]. Figure 3 gives an overview of the possible threat scenarios for different components of IoT/IIoT systems.

In Section 4 we define IoT specific properties to extend the standard threat model approach. Threats are divided into six categories, which are defined in the STRIDE model [21]:

- **Spoofing identity.** A user or service illegally accesses and uses other authentication information to gain illegitimate access to a system or data.

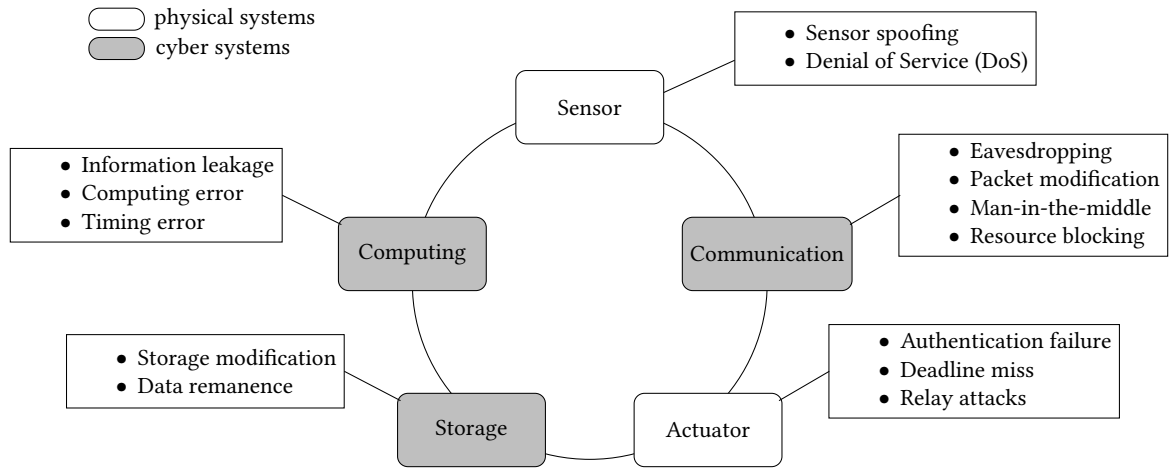

Figure 3: Example attacks on IIoT systems.

- **Tampering with data.** Data tampering occurs when data is malicious modified. This includes data at rest, data in use as well as data in transit.
- **Repudiation** This means that an entity may plausibly deny an action that it has taken. Countering these threats usually requires a combination of authentication, authorization and logging, ideally in a cryptographically secured way.
- **Information disclosure.** Refers to any information exposed to unauthorised users.
- **Denial of service (DoS).** DoS attacks deny services availability to valid users.
- **Elevation of privilege.** These threats occur when unprivileged users gain privileged access and, thus, are able to compromise an entire system.

## 3.2 Security Analysis

In security analysis and cryptanalysis, an information system is studied in order to find hidden or unknown aspects of the system to identify possible weaknesses. Often this mathematical analysis targets detailed aspects of the underlying encryption schemes, protocols, and hardware used in IT system architectures.

The importance of a detailed security analysis of the primitives of encryption schemes and security protocols are highlighted in the case of many recent attacks on IoT systems [47]. Recently, also many international standards and even well studied primitives have been broken. This include cryptographic modes of operation (i.e. OCB2 [22] which was standardised as ISO19772 [23]), the standard hash algorithm SHA-1 [34, 54], and also many block ciphers [4, 20].

While many of this detailed analysis of encryption schemes and protocols are done manually, in recent years the trend to automate parts in this security analysis emerged. Nowadays, many optimisation tools are used to automatically analyse encryption schemes (i.e. using mixed integer linear programming [40], boolean satisfiability solver and satisfiability modular theories solver [5], as well as tools based on constraint programming [55]) or security protocols (e.g. in the symbolic analysis of TLS 1.3 [14]). In an approach to automate the analysis of cryptographic primitives and security protocols, detailed information about the protocol version (e.g. TLS 1.3), the cipher suite (e.g. AES-GCM), block length and key lengths (e.g. 128-bit) are required.

## 3.3 Penetration Testing

Penetration testing is the process of testing computer systems as well as human resources (social engineering) to identify security threats and possible vulnerabilities. These tests can be performed from the *inside* and from the *outside* of the systems under test to ensure that all possible options of an attacker are covered. The aim of each penetration test is to specify guidelines and recommendations which address the identified issues.

*3.3.1 Attack Procedure.* It is essential to define the preconditions of the testing environment as well as the used attack procedure. The preconditions include organisational measures that usually consist of a contract that defines the scope of penetration test, the schedule, people and their availability as well as the reporting procedure. For the attack procedures, we can define three approaches based on the information that an attacker has.

- **Black-box testing:** Only basic information about the target system/network is available, or no information is available.
- **Grey-box testing:** Limited information is available to the attacker (i.e. this can be knowledge about the target network, number of hosts, network infrastructure, ...).
- **White-box testing:** Detailed background information is available to the attacker. Moreover, system information is available (i.e. the attacker has access to the source code, has a detailed overview of the network architecture, ...)

While the black box approach is the most realistic, as an attacker typically accesses the target system from an outside network (e.g. the Internet), an attacker however can also gain access within a network (i.e. by getting physical access). In that case, an attacker can scan the network and obtain valuable information that can later be used in an attack.

*3.3.2 Penetration Testing Phases.* The process of penetration tests can be simplified into the following phases [1, 6]:

- **Reconnaissance:** In this initial phase, information about the target system, software and users is gathered. This information can later be used to attack the target system.
- **Scanning:** In this phase, technical tools are used to further the attacker's knowledge about the target system.
- **Gaining Access:** In this phase, the attacker actively tries to exploit the systems by using the information gathered in the previous phases.
- **Maintaining Access:** An attacker is required to persistently be able to access the target system to gather as much data as possible.
- **Covering Tracks:** An attacker must erase any traces, such as log files, in order to remain anonymous.

*3.3.3 Penetration Testing Tools.* This section gives a short overview about the most common penetration testing tools (Note, all of them are available in Kali Linux [45]) and their usage in certain testing phases.

- **Nmap** is a free open source network scanner which is used within the reconnaissance and scanning phases. It scans single IPs as well as IP ranges and provides useful information like the used operating system, identified services and open ports.
- **Metasploit** is an open source framework which is used for information gathering as well as for executing exploits. The tool comes with a set of pre-installed exploit and auxiliary scripts.
- **Sqlmap** is an open source penetration testing tool which is used to automate the process of exploiting SQL databases. It also can be used for detecting the used database and version. Therefore, the tool can be used within the reconnaissance as well as within the gaining access phase.
- **Subfinder** is an open source subdomain discovery tool.
- **DMitry** is an information gathering tool which can be used to obtain *whois* information, uptime information, email addresses and subdomains. Furthermore, the tool can also be used to perform port scans.
- **Burp Suite** is a powerful platform for security testing of web applications. It can be used to scan automatically for common web vulnerabilities but also offers advanced manual testing techniques to enhance each penetration testing phase.

*3.3.4 Required Input Parameters for Penetration Testing Tools.* We analysed several penetration testing tools (most of them are freely available in Kali Linux [45]) regarding the input requirements they need, to optimally obtain as much information as possible from a target, as well as the information needed to find targeted vulnerabilities and exploits. We arrange the input requirements for the penetration testing tool according to the penetration testing phases as defined above. Table 2 gives an overview of the necessary input information for penetration testing tools. We just list the first three penetration testing phases, as they can easily be automated once there are publicly known exploits available. Using more detailed exploits and maintaining access to a target, often requires user interaction.

Table 2: Penetration testing tool requirements.

| Pentest Phase | Input Requirements |
| --- | --- |
| Reconnaissance | IP address, host names, network addresses, hardware interfaces, ... |
| Scanning | Web URL's IP addresses, network interfaces, port numbers, operating system information, ... |
| Gaining Access | Operating system information, software versions, protocol versions, ... |

*3.3.5 On Automating Security Assurance Tools.* Automating security assurance tools is often cumbersome, as most of the tools do not provide software API's or standardised output formats. Therefore, often security auditors and penetration testers are needed to manually interpret the results and act based on the outputs of the tools. Yet, some of the penetration testing tools, like nmap[2] and metasploit[3], provide an API, store output in XML format, or provide simple scripting languages to automate scanning and testing.

# 4 REQUIREMENTS AND RECOMMENDATIONS FOR IOT/IIOT MODELS

Security testing is the process to validate and verify that a system under test meets its security requirements. As already outlined in Section 2, a model-based approach can be used to abstract a real-world IoT/IIoT system. Model-based security testing is discussed in more detail in [50].

In large industrial IoT systems, the complexity of the system is often too high for manual testing. Moreover, to improve effectiveness, efficiency, quality and to be able to repeat a testing process as often as required, automated testing systems are needed. However, to automate security testing of IoT/IIoT systems, the IoT/IIoT models that abstract the real-world systems under test have to be as accurate as possible.

We give a non-exhaustive list of security requirements that models of IoT/IIoT components should include, to optimise the automated information gathering phase of the security assurance tools (as discussed in Section 3). In general, the more information can be abstracted from the real-world into a model, the more information can then be used in the attacks.

## 4.1 Limitations and Restrictions

Depending on the size of a IoT system a detailed security analysis and detailed penetration tests can quickly exhaust the available resources. Therefore, we have to limit the level of detail of the security analysis and take some assumptions into consideration. In general, we assume that the cryptographic primitives are based on well-studied standards and therefore we do not consider detailed attacks on the building blocks (i.e. such as block ciphers, cryptographic modes). Yet, this does not include misconfiguration of cipher suites, as these belong to the most common errors. Moreover, not all steps in security tests can be automated. Therefore, some steps of security tests have to be managed by experienced security auditors and cryptographers.

---
[2]https://nmap.org/book/nse-api.html
[3]https://metasploit.help.rapid7.com/docs/rest-api

Table 3: Desired parameters of IoT/IIoT components. *SD&S:* Smart Devices and Sensors, *N&P:* Networks and Protocols, *G:* Gateways, *C&S:* Cloud and Servers, *SS&BS:* Smart Services and Backend Systems, *UI:* User Interfaces. The priority for each parameter is indicated with colour (red = high, yellow = medium, green = low).

| Parameter | Description | IoT/IIoT Components | | | | | |
|---|---|---|---|---|---|---|---|
| | | SD&S | N&P | G | C&S | SS&BS | UI |
| *Network Properties* | | | | | | | |
| Hardware interface | Indicates all available interfaces (i.e. network, USB). | ✓ | | ✓ | ✓ | ✓ | |
| Connection type | Indicates if the network is a physical network (Ethernet) or if the network is wireless (WIFI, Bluetooth, NFC, ...). | ✓ | ✓ | ✓ | ✓ | ✓ | |
| IP address | Indicates the IPv4/IPv6 addresses. | ✓ | | ✓ | ✓ | ✓ | |
| MAC address | Indicates the MAC addresses. | ✓ | | ✓ | ✓ | ✓ | |
| Network protocols | Lists all supported network protocols. | ✓ | | ✓ | ✓ | ✓ | |
| Protocol version | Describes the protocol version used (i.e. TLS 1.3). | ✓ | ✓ | ✓ | | | |
| Pairing process | Indicates how the device connects to other devices. | ✓ | | ✓ | ✓ | ✓ | |
| *Hardware Properties* | | | | | | | |
| Secure key store | Indicates if a secure key store is available (i.e. Secure Element, Hardware Security Module, or Trusted Platform Module). | ✓ | | ✓ | ✓ | ✓ | |
| Data storage | Indicates if data is stored internally. | ✓ | | ✓ | ✓ | ✓ | |
| Power consumption | Indicates the power consumption. | ✓ | | | | | |
| Electromagnetic emission | Indicates the electromagnetic emission. | ✓ | | | | | |
| *Software & Operating System Properties* | | | | | | | |
| Operating system | Indicates the operating system version (i.e. Windows 10). | ✓ | | ✓ | ✓ | ✓ | |
| Firmware version | Indicates the firmware version. | ✓ | | ✓ | | | |
| Software APIs | Lists the available APIs. | ✓ | | | ✓ | ✓ | ✓ |
| Software versions | Lists the other software versions running on the device. | ✓ | | ✓ | ✓ | ✓ | |
| Interfaces | Lists the available interfaces that can be used to connect to a IoT system (i.e. graphical UIs, command line UIs). | ✓ | | | ✓ | ✓ | ✓ |
| Administration | Indicates how the system is maintained (remote access). | ✓ | ✓ | ✓ | ✓ | ✓ | ✓ |
| Update process | Indicates how software/firmware updates are received (i.e. over the air). | ✓ | ✓ | ✓ | ✓ | ✓ | ✓ |
| Reset functionality | Indicates how the device can be reset to an initial setting. | ✓ | | ✓ | ✓ | ✓ | |
| Shared resources | Indicates if resources are shared with other users/processes. | | | ✓ | ✓ | ✓ | |
| *Security Properties* | | | | | | | |
| Encryption | Indicates if the channel is end-to-end encrypted. | | ✓ | | | | |
| Data integrity | Indicates if the data integrity of the connection is ensured. | | ✓ | | | | |
| Authentication | Indicates if authentication is required. | | ✓ | | | | ✓ |
| Input sanitization | Indicates if the user inputs are checked before forwarded to backend systems. | | ✓ | | | | ✓ |
| *Performance Properties* | | | | | | | |
| Bandwidth | Indicates the maximum rate of network packets that are sent in a time interval (measured in bits/second). | | ✓ | ✓ | ✓ | | |
| Throughput | Indicates the actual rate of network packets that are sent in a time interval (measured in bits/second). | | ✓ | ✓ | ✓ | | |
| Latency | Indicates the delay between a data packet sent from the source until it is received at the destination. | | ✓ | ✓ | ✓ | | |
| Error rate | Indicates the number of corrupted bits as a percentage of the total sent bits. | | ✓ | ✓ | ✓ | | |

## 4.2 Required Properties of IoT/IIoT Models for Safe and Secure IoT/IIoT Systems

In the following, we list desired parameters that should be considered in IoT/IIoT models. If some parameters are unknown in the initial modelling phase, they can be added/updated later during an information gathering phase, which is part of penetration testing, or be specified when a more detailed threat model is generated. Table 3 lists the desired properties for the below mentioned IoT/IIoT components.

*4.2.1 Smart Devices and Sensors.* Smart devices and sensors that are connected to networks are potential threats, as they can have security vulnerabilities due to outdated operating systems, insecure software versions, open network ports, unsecured hardware interfaces, and many more security issues [26].

*4.2.2 Networks and Protocols.* IoT networks are used to interconnect several IoT devices and sensors with back-end or cloud systems via the Internet. Often different types of networks are used, also with many different protocols. Potential threats in IoT networks can occur, if outdated protocols are used that have security vulnerabilities [8, 25, 27], or if different networks have different security requirements.

*4.2.3 Gateways.* IoT gateways are physical devices or software programs that serve as a connection point between smart devices/sensors and servers in a cloud. Those gateways are used to perform protocol translation, data processing, data storage and filtering. Potential threats can occur in the protocol translation when different security requirements are set or unsecured network interfaces, open hardware ports or vulnerable software is used.

*4.2.4 Cloud/Servers.* Cloud-based services are often used in IoT systems for infrastructure, data storage, data processing and analytics. Usually these services are provided by third parties. Therefore, potential threats can occur if data is not encrypted, or an adversary has access to any servers in the particular cloud.

*4.2.5 Smart Service and Back-end Systems.* Often IIoT devices and sensors in smart factories produce huge amounts of data that are first filtered and pre-processed in a cloud system, and then further processed and stored in back-end systems within a company network. Potential threats for those smart services and back-end systems can occur from denial of service (DoS) attacks, as well as through open network ports, outdated software, misconfigured hardware/software and other security issues.

*4.2.6 User Interfaces.* User interfaces are the accessible parts of IoT systems, that are normally controlled by human users. Potential threats can occur as interfaces as well as users themselves are a relatively easy link to exploit. This can occur from misconfiguration in user interfaces, weak passwords, social engineering [28], phishing [17] and software vulnerabilities.

## 5 CONCLUSION

In this paper we analyse the current state of security assurance methodologies such as threat modelling, security analysis and penetration testing in the context of IoT/IIoT systems. After identifying shortcomings in the manual testing strategies that are sufficient in small scale computer networks, we pinpoint the need for automated security analysis in large scale IoT/IIoT networks.

In that context, we first classify the typical components of IoT/IIoT systems. By analysing the most commonly used security testing tools, we identify a list of required input parameters and metadata that needs to be available at the time of a detailed security analysis. Therefore, we define detailed requirement and recommendations for metadata that needs to be present in IoT/IIoT system models, which are found along the software development process. Using this metadata information, the security assurance process can efficiently be automated, making detailed security testing possible in even large scale IoT/IIoT networks.

## ACKNOWLEDGMENTS

This work was partly supported by the Austrian Research Promotion Agency (FFG) within the ICT of the future grants program, grant nb. 863129 (project IoT4CPS), of the Federal Ministry for Transport, Innovation and Technology (BMVIT).

## REFERENCES

[1] Farkhod Alisherov A and Feruza Sattarova Y. 2009. Methodology for Penetration Testing. *International Journal of Grid and Distributed Computing* (2009), 43–50.
[2] ZigBee Alliance. 2012. ZigBee Specification, 053474r20.
[3] N. A. Almubairik and G. Wills. 2016. Automated penetration testing based on a threat model. In *2016 11th International Conference for Internet Technology and Secured Transactions (ICITST)*. 413–414. https://doi.org/10.1109/ICITST.2016.7856742
[4] Ralph Ankele, Christoph Dobraunig, Jian Guo, Eran Lambooij, Gregor Leander, and Yosuke Todo. 2019. Zero-Correlation Attacks on Tweakable Block Ciphers with Linear Tweakey Expansion. *IACR Transactions on Symmetric Cryptology* 2019, 1 (Mar. 2019), 192–235. https://doi.org/10.13154/tosc.v2019.i1.192-235
[5] Ralph Ankele and Stefan Kölbl. 2019. Mind the Gap - A Closer Look at the Security of Block Ciphers against Differential Cryptanalysis. In *Selected Areas in Cryptography – SAC 2018*, Carlos Cid and Michael J. Jacobson Jr. (Eds.). Springer International Publishing, Cham, 163–190.
[6] B. Arkin, S. Stender, and G. McGraw. 2005. Software penetration testing. *IEEE Security Privacy* 3, 1 (Jan 2005), 84–87. https://doi.org/10.1109/MSP.2005.23
[7] Andrew Banks and Rahul Gupta. 2014. MQTT Version 3.1.1. http://docs.oasis-open.org/mqtt/mqtt/v3.1.1/os/mqtt-v3.1.1-os.html
[8] B. Beurdouche, K. Bhargavan, A. Delignat-Lavaud, C. Fournet, M. Kohlweiss, A. Pironti, P. Strub, and J. Zinzindohoue. 2015. A Messy State of the Union: Taming the Composite State Machines of TLS. In *2015 IEEE Symposium on Security and Privacy (SP)*. IEEE Computer Society, Los Alamitos, CA, USA, 535–552. https://doi.org/10.1109/SP.2015.39
[9] John Biggs. 2016. Hackers release source code for a powerful DDoS app called Mirai. https://techcrunch.com/2016/10/10/hackers-release-source-code-for-a-powerful-ddos-app-called-mirai/?guccounter=1.
[10] Hugh Boyes, Bil Hallaq, Joe Cunningham, and Tim Watson. 2018. The industrial internet of things (IIoT): An analysis framework. *Computers in Industry* 101 (2018), 1 – 12. https://doi.org/10.1016/j.compind.2018.04.015
[11] Malte Brettel, Niklas Friederichsen, Michael Keller, and Marius Rosenberg. 2014. How Virtualization, Decentralization and Network Building Change the Manufacturing Landscape: An Industry 4.0 Perspective. *International journal of mechanical, aerospace, industrial and mechatronics engineering* 8, 1 (2014), 37–44. https://publications.rwth-aachen.de/record/465283
[12] Ge Chu and Alexei Lisitsa. 2019. Penetration Testing for Internet of Things and Its Automation. https://doi.org/10.1109/HPCC/SmartCity/DSS.2018.00244
[13] LoRa Alliance Technical Committee. 2017. LoRaWAN[TM] 1.1 Specification.
[14] Cas Cremers, Marko Horvat, Jonathan Hoyland, Sam Scott, and Thyla van der Merwe. 2017. A Comprehensive Symbolic Analysis of TLS 1.3. In *Proceedings of the 2017 ACM SIGSAC Conference on Computer and Communications Security (CCS '17)*. ACM, New York, NY, USA, 1773–1788. https://doi.org/10.1145/3133956.3134063

[15] S. Deering and R. Hinden. 1998. *RFC 2460 Internet Protocol, Version 6 (IPv6) Specification.* Internet Engineering Task Force. http://tools.ietf.org/html/rfc2460
[16] V. A. Desnitsky, I. V. Kotenko, and S. B. Nogin. 2015. Detection of anomalies in data for monitoring of security components in the Internet of Things. In *2015 XVIII International Conference on Soft Computing and Measurements (SCM).* 189–192. https://doi.org/10.1109/SCM.2015.7190452
[17] Rachna Dhamija, J Doug Tygar, and Marti Hearst. 2006. Why phishing works. In *Proceedings of the SIGCHI conference on Human Factors in computing systems.* ACM, 581–590.
[18] D. Dolev and A. Yao. 1983. On the Security of Public Key Protocols. *IEEE Trans. Inf. Theor.* 29, 2 (March 1983), 198–208. https://doi.org/10.1109/TIT.1983.1056650
[19] S. Earley. 2015. Analytics, Machine Learning, and the Internet of Things. *IT Professional* 17, 1 (Jan 2015), 10–13. https://doi.org/10.1109/MITP.2015.3
[20] Viet Tung Hoang, David Miller, and Ni Trieu. 2019. Attacks Only Get Better: How to Break FF3 on Large Domains. Cryptology ePrint Archive, Report 2019/244. https://eprint.iacr.org/2019/244.
[21] Michael Howard and David LeBlanc. 2003. *Writing secure code.* Pearson Education.
[22] Akiko Inoue, Tetsu Iwata, Kazuhiko Minematsu, and Bertram Poettering. 2019. Cryptanalysis of OCB2: Attacks on Authenticity and Confidentiality. Cryptology ePrint Archive, Report 2019/311. https://eprint.iacr.org/2019/311.
[23] ISO/IEC 19772:2009 2009. *Information technology – Security techniques – Authenticated encryption.* Standard. International Organization for Standardization, Geneva, CH.
[24] Tanner Johnson. 2018. Growing Cybersecurity Concerns Within the Industrial IoT (IIoT). https://technology.ihs.com/607003/growing-cybersecurity-concerns-within-the-industrial-iot-iiot Accessed: 2019-05-15.
[25] Antti Karjalainen, Riku Hietamäki, Matti Kamunen, and Neel Mehta. 2013. OpenSSL 'Heartbleed' vulnerability CVE-2014-0160. Available from MITRE, CVE-ID CVE-2014-0160.. http://cve.mitre.org/cgi-bin/cvename.cgi?name=CVE-2014-0160
[26] C. Kolias, G. Kambourakis, A. Stavrou, and J. Voas. 2017. DDoS in the IoT: Mirai and Other Botnets. *Computer* 50, 7 (2017), 80–84. https://doi.org/10.1109/MC.2017.201
[27] Bodo Moller, Thai Duong, Krzysztof Kotowicz. 2014. *This POODLE Bites: Exploiting The SSL 3.0 Fallback.* https://www.openssl.org/~bodo/ssl-poodle.pdf
[28] Katharina Krombholz, Heidelinde Hobel, Markus Huber, and Edgar Weippl. 2015. Advanced social engineering attacks. *Journal of Information Security and applications* 22 (2015), 113–122.
[29] Silicon Labs. 2019. Z-Wave Plus Device Type v2 Specification.
[30] P. Lade, R. Ghosh, and S. Srinivasan. 2017. Manufacturing Analytics and Industrial Internet of Things. *IEEE Intelligent Systems* 32, 3 (May 2017), 74–79. https://doi.org/10.1109/MIS.2017.49
[31] Selena Larson. 2017. FDA confirms that St. Jude's cardiac devices can be hacked. https://money.cnn.com/2017/01/09/technology/fda-st-jude-cardiac-hack/.
[32] E. A. Lee. 2008. Cyber Physical Systems: Design Challenges. In *2008 11th IEEE International Symposium on Object and Component-Oriented Real-Time Distributed Computing (ISORC).* 363–369. https://doi.org/10.1109/ISORC.2008.25
[33] Jay Lee, Behrad Bagheri, and Hung-An Kao. 2015. A Cyber-Physical Systems architecture for Industry 4.0-based manufacturing systems. *Manufacturing Letters* 3 (2015), 18 – 23. https://doi.org/10.1016/j.mfglet.2014.12.001
[34] Gaëtan Leurent and Thomas Peyrin. 2019. From Collisions to Chosen-Prefix Collisions - Application to Full SHA-1. Cryptology ePrint Archive, Report 2019/459. https://eprint.iacr.org/2019/459.
[35] S. Lipner. 2004. The trustworthy computing security development lifecycle. In *20th Annual Computer Security Applications Conference.* 2–13. https://doi.org/10.1109/CSAC.2004.41
[36] Duško Lukač. 2015. The fourth ICT-based industrial revolution "Industry 4.0" – HMI and the case of CAE/CAD innovation with EPLAN P8. 835–838. https://doi.org/10.1109/TELFOR.2015.7377595
[37] Aaron Marback, Hyunsook Do, Ke He, Samuel Kondamarri, and Dianxiang Xu. 2013. A threat model – based approach to security testing. *Software: Practice and Experience* 43 (02 2013). https://doi.org/10.1002/spe.2111
[38] Stefan Marksteiner, Víctor Juan Expósito Jiménez, Heribert Vallant, and Herwig Zeiner. 2017. An overview of wireless IoT protocol security in the smart home domain. In *Proceedings of 2017 Internet of Things Business Models, Users, and Networks Conference (CTTE).* 1–8. https://doi.org/10.1109/CTTE.2017.8260940
[39] Charlie Miller and Chris Valasek. 2015. Remote Exploitation of an Unaltered Passenger Vehicle. https://www.blackhat.com/us-15/briefings.html#remote-exploitation-of-an-unaltered-passenger-vehicle.
[40] Nicky Mouha, Qingju Wang, Dawu Gu, and Bart Preneel. 2012. Differential and Linear Cryptanalysis Using Mixed-Integer Linear Programming. In *Information Security and Cryptology,* Chuan-Kun Wu, Moti Yung, and Dongdai Lin (Eds.). Springer Berlin Heidelberg, Berlin, Heidelberg, 57–76.
[41] JaeKwan Park, YongSuk Suh, and Cheol Park. 2016. Implementation of cyber security for safety systems of nuclear facilities. *Progress in Nuclear Energy* 88 (2016), 88 – 94. https://doi.org/10.1016/j.pnucene.2015.12.009
[42] R. Poovendran. 2010. Cyber-Physical Systems: Close Encounters Between Two Parallel Worlds [Point of View]. *Proc. IEEE* 98, 8 (Aug 2010), 1363–1366. https://doi.org/10.1109/JPROC.2010.2050377
[43] Bluetooth SIG Proprietary. 2019. Bluetooth Core Specification V5.1.
[44] R. Rajkumar, I. Lee, L. Sha, and J. Stankovic. 2010. Cyber-physical systems: The next computing revolution. In *Design Automation Conference.* 731–736. https://doi.org/10.1145/1837274.1837461
[45] Vivek Ramachandran and Cameron Buchanan. 2015. *Kali Linux: Wireless Penetration Testing Beginner's Guide.* Packt Publishing.
[46] Shahid Raza, Linus Wallgren, and Thiemo Voigt. 2013. SVELTE: Real-time intrusion detection in the Internet of Things. *Ad Hoc Networks* 11, 8 (2013), 2661 – 2674. https://doi.org/10.1016/j.adhoc.2013.04.014
[47] E. Ronen, A. Shamir, A. Weingarten, and C. O'Flynn. 2017. IoT Goes Nuclear: Creating a ZigBee Chain Reaction. In *2017 IEEE Symposium on Security and Privacy (SP).* 195–212. https://doi.org/10.1109/SP.2017.14
[48] Vinay Sachidananda, Shachar Siboni, Asaf Shabtai, Jinghui Toh, Suhas Bhairav, and Yuval Elovici. 2017. Let the Cat Out of the Bag: A Holistic Approach Towards Security Analysis of the Internet of Things. In *Proceedings of the 3rd ACM International Workshop on IoT Privacy, Trust, and Security (IoTPTS '17).* ACM, New York, NY, USA, 3–10. https://doi.org/10.1145/3055245.3055251
[49] A. Sajid, H. Abbas, and K. Saleem. 2016. Cloud-Assisted IoT-Based SCADA Systems Security: A Review of the State of the Art and Future Challenges. *IEEE Access* 4 (2016), 1375–1384. https://doi.org/10.1109/ACCESS.2016.2549047
[50] Ina Schieferdecker, Jürgen Großmann, and Martin Schneider. 2012. Model-Based Security Testing. *Electronic Proceedings in Theoretical Computer Science* 80 (02 2012). https://doi.org/10.4204/EPTCS.80.1
[51] Wentao Shang, Yingdi Yu, Ralph Droms, and Lixia Zhang. 2016. Challenges in IoT networking via TCP/IP architecture. *NDN, Technical Report NDN-0038* (2016).
[52] Adam Shostack. 2014. *Threat modeling: Designing for security.* John Wiley & Sons.
[53] Statista. 2019. Internet of Things (IoT) connected devices installed base worldwide from 2015 to 2025 (in billions). https://www.statista.com/statistics/471264/iot-number-of-connected-devices-worldwide/. Accessed: 2019-05-06.
[54] Marc Stevens, Elie Bursztein, Pierre Karpman, Ange Albertini, and Yarik Markov. 2017. The First Collision for Full SHA-1. In *Advances in Cryptology – CRYPTO 2017,* Jonathan Katz and Hovav Shacham (Eds.). Springer International Publishing, Cham, 570–596.
[55] Siwei Sun, David Gerault, Pascal Lafourcade, Qianqian Yang, Yosuke Todo, Kexin Qiao, and Lei Hu. 2017. Analysis of AES, SKINNY, and Others with Constraint Programming. *IACR Transactions on Symmetric Cryptology* 2017, 1 (Mar. 2017), 281–306. https://doi.org/10.13154/tosc.v2017.i1.281-306
[56] S. Türpe. 2017. The Trouble with Security Requirements. In *2017 IEEE 25th International Requirements Engineering Conference (RE).* 122–133. https://doi.org/10.1109/RE.2017.13
[57] Shiyong Wang, Jiafu Wan, Di Li, and Chunhua Zhang. 2016. Implementing smart factory of industrie 4.0: an outlook. *International Journal of Distributed Sensor Networks* 12, 1 (2016), 3159805.
[58] Dianxiang Xu, Manghui Tu, Michael Sanford, Lijo Thomas, Daniel Woodraska, and Weifeng Xu. 2012. Automated Security Test Generation with Formal Threat Models. *IEEE Transactions on Dependable and Secure Computing - TDSC* 9 (07 2012), 526–540. https://doi.org/10.1109/TDSC.2012.24
[59] Jonathan Yung, Hervé Debar, and Louis Granboulan. 2017. Security Issues and Mitigation in Ethernet POWERLINK. In *Security of Industrial Control Systems and Cyber-Physical Systems,* Nora Cuppens-Boulahia, Costas Lambrinoudakis, Frédéric Cuppens, and Sokratis Katsikas (Eds.). Springer International Publishing, Cham, 87–102.
[60] Bruno Bogaz Zarpelão, Rodrigo Sanches Miani, Cláudio Toshio Kawakani, and Sean Carlisto de Alvarenga. 2017. A survey of intrusion detection in Internet of Things. *Journal of Network and Computer Applications* 84 (2017), 25 – 37. https://doi.org/10.1016/j.jnca.2017.02.009